%% file: main.tex
\documentclass[notitlepage,superscriptaddress,pra,twocolumn,nofootinbib]{revtex4-1}
\usepackage{siunitx}
\usepackage{amsmath}
\usepackage{amsfonts}
\usepackage{amssymb}
\usepackage{graphicx}
\usepackage{sidecap}
\usepackage{braket}
\usepackage{blindtext}
\usepackage{subfigure}
\usepackage{lipsum}
\usepackage{fullpage}
\usepackage{appendix}
\usepackage{subfigure}
\usepackage[dvipsnames]{xcolor}
\usepackage{physics}
\usepackage{hyperref}
\usepackage{amsthm}
\newcommand{\hc}{\text{H.c.}}
\newcommand{\ex}[1]{\langle #1 \rangle}
\hypersetup{
	colorlinks=true,       % false: boxed links; true: colored links
	linkcolor=VioletRed,  % color of internal links (change box color with linkbordercolor)
    citecolor=JungleGreen,        % color of links to bibliography
    filecolor=Orchid,      % color of file links
    urlcolor=black           % color of external links
}

\begin{document}
\title{Tunable frequency-bin multi-mode squeezed states of light}
\author{Christian Drago}
\email{christian.drago@mail.utoronto.ca}
\affiliation{Department of Physics and Astronomy, University of Waterloo, 200 University Avenue West, Waterloo, Ontario, Canada N2L 3G1}
\affiliation{Department of Physics, University of Toronto, 60 St. George Street, Toronto, Ontario, Canada, M5S 1A7}
\author{Agata M. Bra\'nczyk}
\affiliation{Perimeter Institute for Theoretical Physics, Waterloo, Ontario, N2L 2Y5, Canada}

\begin{abstract}
Squeezed states are a versatile class of quantum states with applications ranging from quantum computing to high-precision detection. We propose a method for generating tunable squeezed states of light with multiple modes encoded in frequency bins. Our method uses custom-engineered spontaneous parametric downconversion pumped by a pulse-shaped pump field. The multi-mode squeezed states are generated in a single pass and can be tuned in real time by adjusting the properties of the pump field.  Exploring new
quantum states of light, encoded in new degrees of
freedom, can be a fruitful path toward discovering new quantum applications. 
\end{abstract}

\maketitle

\section{Introduction}\label{sec:Introduction}
Quantum light is a key ingredient in emerging quantum technologies such as quantum communication \cite{milburn1999quantum}, quantum cryptography \cite{hillery2000quantum}, quantum computing \cite{arrazola2021quantum}, quantum imaging \cite{bennink2004quantum} and quantum metrology \cite{schnabel2010quantum}. Its broad application results from the wide variety of available degrees of freedom (e.g. polarization, frequency, temporal mode), which in turn can encode a wide variety of quantum states (e.g. qubits, Fock states, cat states). Exploring new quantum states of light, encoded in new degrees of freedom, can thus be a fruitful path toward discovering new quantum applications.   

A versatile class of quantum states are know as \emph{squeezed vacuum states}  \cite{lvovsky2014squeezed}, or simply \emph{squeezed states}\footnote{For a brief review of squeezed states, refer to Appendix \ref{sec:Background}.}. They are typically classified by the number of modes they populate, where different kinds of squeezed states are better suited for different applications. \emph{Single-mode} squeezed states have reduced quantum noise in one degree of freedom, making them most useful for quantum cryptography \cite{hillery2000quantum} and quantum metrology \cite{schnabel2010quantum}. \emph{Two-mode} squeezed states, on the other hand, posses entanglement between the modes of the electromagnetic field, and can be used for quantum teleportation \cite{milburn1999quantum}. Squeezed states can also be distributed across \emph{multiple} modes. These \emph{multi-mode}  squeezed states can be used to generalize various quantum information protocols, e.g. multi-parameter quantum metrology \cite{gessner2018sensitivity}, multi-channel quantum imaging \cite{sokolov2004squeezed}, multi-partite teleportation \cite{lian2007continuous} or be used for quantum computation \cite{fabre2020generation}.

The generalization of single- and two-mode squeezed states was first proposed by Yeoman and Barnett in 1993 \cite{Yeoman1993TwomodeSG}. They considered states (originally called \emph{two-mode squeezed gaussons}) that had both single- and two-mode squeezing properties and proposed a method to generate them with two single-mode squeezed states and a beam splitter. In 2000 this method was generalized to include multi-mode squeezed states by Van Loock and Braunstein \cite{van2000multipartite} who proposed using a sequence of $N$ beam splitters to prepare a multimode squeezed state entangled across $N$ spatial modes. This method requires interferometric stabilization of the optical paths and indistinguishability between frequency and polarization modes \cite{elezov2018active}. To overcome these challenges, another approach was proposed \cite{pfister2004multipartite, menicucci2007ultracompact} in which an optical parametric oscillator (OPO) and a single periodically poled ferroelectric crystal is used to generate $N$-mode entangled states between the cavity modes of the OPO. This method has been further expanded upon by multiplexing the light in time allowing an unlimited number of entangled modes \cite{alexander2016one,zhu2021hypercubic}. 
Further, one can encode squeezed state modes into Schmidt modes as was done in \cite{ansari2018tailoring}. These newer methods are much more compact and stable, and benefit from the intrinsic compatibility of frequency and time encodings with waveguides and fiber transmission, and we expand on them here. 

We introduce a method for generating tunable multi-mode squeezed states of light encoded in frequency bins. Our proposal builds on the idea that frequency-bin entanglement can be generated by domain-engineered down-conversion, which was recently demonstrated by  Morrison \emph{et al.} \cite{morrison2022frequency}. We show how the addition of a frequency-shaped pump can yield grid states similar to those introduced by Fabre \emph{et al.} \cite{fabre2020generation}, but with more control over the peaks, and without the need for a cavity. We further show how, with the addition of frequency fitering, one can create a squeezed state that can be tuned from single-mode to two-mode in real time. 

Our method for generating multimode squeezed states differs from previous proposals in several ways: (i) unlike \cite{pfister2004multipartite, menicucci2007ultracompact, alexander2016one,zhu2021hypercubic,fabre2020generation}, our method does not require a cavity to generate multiple modes, (ii) unlike \cite{Yeoman1993TwomodeSG,van2000multipartite}, our method does not require increasingly more beam splitters to generate more modes, and (iii) unlike the Schmidt modes in \cite{ansari2018tailoring}, which are difficult to distinguish experimentally due to their complex spectral shape, our modes are encoded in discrete frequency bins and can be distinguished by their central frequencies. Furthermore, our method allows the squeezing parameters of the state to be tuned in real time. Depending on the application, these properties may make this method advantageous over other methods.

\section{Frequency bin encoding}\label{sec:Pulsed-mode encoding}
A desirable property of quantum light is localization in space and time, which makes it possible to deliver wavepackets of quantum light in a clocked manner \cite{lloyd1993programming}. A Gaussian-like frequency encoding satisfies these properties. As with their CW counterparts \cite{pfister2004multipartite, menicucci2007ultracompact}, such an encoding is intrinsically compatible with waveguides and fiber transmission.

To encode our modes we define a multi-frequency field operator: 
\begin{align}
    \label{eq:A_n}
     A_{n} = \int d\omega G^*_n(\omega)  a(\omega)\,,
\end{align}
where  $\{G_n(\omega)\}$ describe non-overlapping Gaussian-like pulses, such that to good approximation
\begin{align}
    \int d\omega G_n(\omega) G_m(\omega) = \delta_{nm}\,,
\end{align}
and thus
\begin{align}
    [  A_{n},  A^{\dagger}_{m}] = \delta_{nm}\,.
\end{align}
In terms of these operators, the multi-mode squeezed state takes the form
\begin{align}
    \begin{split}
        \label{eq:mmsv An}
        \ket{\psi}  =  e^{\frac{1}{2} \sum\limits_{n,m} \gamma_{nm}  {A}_{n}^{\dagger}  {A}_{m}^{\dagger}
        -\hc}\ket{\text{vac}}\,,
    \end{split}
\end{align}
which can be rewritten as
\begin{align}
        \ket{\psi} = e^{
        \frac{1}{2}\int d\omega_i d\omega_s h(\omega_s,\omega_i)  {a}^{\dagger}(\omega_i)  {a}^{\dagger}(\omega_s) - H.c. }\ket{\text{vac}}\,,
        \label{eq:type0 squ light}
\end{align}
where 
\begin{align}\label{eq:jsa-t1}
    h(\omega_s,\omega_i)=\sum\limits_{n,m} \gamma_{nm} G_n(\omega_i) G_m(\omega_s),
\end{align}
and $\gamma_{nm}$ is in general complex and taken to be symmetric. The state in Eq. \eqref{eq:type0 squ light} could in principle be generated by Type-I spontaneous parameteric downconversion (where both idler and signal modes have the same polarization), \emph{if} one had means of engineering a joint spectral amplitude of the form given by \eqref{eq:jsa-t1}. To date, most joint-spectral engineering methods have been developed for Type-II parameteric downconversion (where both idler and signal have orthogonal polarization), so we will focus on a Type-II setup in this paper. In the next section, we show how to use spectrally engineered Type-II downconversion to generate a state that, once sent through a Mach-Zehnder interferometer with a half-wave-plate (HWP) set to $45^{\circ}$ in one arm, is equal to the multi-mode squeezed state in \eqref{eq:mmsv An}.

\section{Implementation with spontaneous parameteric downconversion}
\label{sec:Implementation with spontaneous parameteric downconversion}
We consider Type-II spontaneous parametric downconversion \cite{couteau2018spontaneous}, under the following conditions: (i) rotating wave approximation, (ii) undepleted pump approximation, (iii) ignoring time-ordering effects, (iv) symmetric group-velocity-matching, and (v) linear phase-mismatch (see Appendix \ref{sec:SPDC} for discussion of the assumptions). In this regime, the state generated by this process is:
\begin{align}\label{eq:type2 squ light}
    \ket{\psi} = e^{i\gamma\int d\omega_i d\omega_s f(\omega_s,\omega_i)  {a}^{\dagger}_{H}(\omega_i)  {a}^{\dagger}_{V}(\omega_s) - H.c.}\ket{\text{vac}}\,,
\end{align}
where  $\gamma$ is the squeezing parameter and
\begin{align}\label{eq:JSAreal}
    f (\omega_s,\omega_i) =\alpha(\omega_s + \omega_i)\phi\left(\omega_s-\omega_i\right),
\end{align}
is the joint spectral amplitude (JSA). The JSA is given by the spectral profile of the pump $\alpha(\omega_s + \omega_i)$ and the normalized phasematching function $\phi\left(\omega_s-\omega_i\right)$ which depends on the properties of the nonlinear material. Following the normalization of the pump and phase-matching function in Appendix \ref{sec:SPDC}, the joint spectral amplitude is normalized according to
\begin{equation}
    \int d\omega_sd\omega_i |f(\omega_s,\omega_i)|^2 = 1.
\end{equation}

\subsection{Designing the joint spectral amplitude}

Our goal is to design the pump-envelope function $\alpha(\omega_s + \omega_i)$ and the phase matching function  $\phi\left(\omega_s-\omega_i\right)$ such that the joint spectral amplitude $f(\omega_s,\omega_i)$ in Eq. \eqref{eq:JSAreal} matches the target joint spectral amplitude $h(\omega_s,\omega_i)$ in Eq. \eqref{eq:jsa-t1}.

To achieve this, the pump envelope function should be prepared as a superposition of Gaussian-like functions centred at different frequencies: 
\begin{align}    \label{eq-generalization-general PUMP}
    \alpha(\omega_s+\omega_i) = \sum_{n=-N}^{N} \frac{a_n}{(4\pi\sigma^2)^{\frac{1}{4}}}e^{-\frac{(\omega_s+\omega_i - \Omega_p + n\delta\omega)^2}{8\sigma^2}},
\end{align}
for some set of dimensionless constants $a_n$ and integers $n$. The pump envelope function  can be customized with various pulse shaping techniques; for a review of pulse shaping in various regimes see \cite{weiner2011ultrafast,froehly1983ii,monmayrant2010newcomer}.
Since the pump function is square normalized, the coefficients satisfy $ \sum_{n}a_n^2 = 1$.

The phase-matching function should also be a superposition of Gaussian-like functions, centred at different frequencies:
\begin{align}    \label{eq:ch-5-general PMF}
    \phi(\omega_s-\omega_i) {=}\hspace{-2mm}\sum_{m=-N}^{N} \frac{b_m}{(4\pi\sigma^2)^{\frac{1}{4}}}e^{\frac{-(\omega_s -\omega_i -(\Omega_s-\Omega_i) -  m\delta\omega)^2}{8\sigma^2}}\,,
\end{align}
where  $b_m$ are some set of dimensionless amplitude coefficients that satisfy $ \sum_{m}b_m^2 = 1$, and $m$ is an integer.  

We will show how to customize the phase-matching function using custom-poling in Section \ref{sec:cust}. In the mean time, we note that if the width of each term in the PMF and pump are the same, and we insert Eqs. \eqref{eq-generalization-general PUMP} and \eqref{eq:ch-5-general PMF} into Eq. \eqref{eq:JSAreal} we yield  the joint spectral amplitude
\begin{align}\label{eq:JSAreal2}
    f (\omega_s,\omega_i) =\frac{1}{\sqrt{2}}\sum_{n,m}a_{n-m}b_{n+m} G_{n}(\omega_s)G_{m}(\omega_i),
\end{align}
where for the remainder of this section, the summations are over $n$ and $m$ such that $n-m = \pm1,\pm2,...,\pm N$, and 
\begin{align}\label{eq:gaussian}
    G_n(\omega_J)= \frac{e^{-\frac{(\omega_J - \Omega_J- \Omega_n)^2}{4\sigma^2}}}{(2\pi\sigma^2)^{\frac{1}{4}}}\,,
\end{align}
with $\Omega_n \equiv n\delta\omega/2$ and $J=s,i$ runs over the two signal and idler frequencies. When absolute-value-squared, these functions yield Gaussian intensity distributions, which are optimal for such decompositions \cite{quesada2018gaussian}. Finally, if $|\Omega_n - \Omega_m|\gg \sigma$, then to good approximation
\begin{align}
    \int d\omega G_n(\omega) G_m(\omega)= \delta_{nm}.
\end{align}

Inserting the decomposition of the JSA in Eq. \eqref{eq:JSAreal2} into the downconverted state in Eq. \eqref{eq:type2 squ light}, defining $\gamma_{nm} \equiv \gamma a_{n-m}b_{n+m}/\sqrt{2}$ and using the definition of the mode $A_n$ in Eq. \eqref{eq:A_n}, yields
\begin{align}\label{eq:polstate}
    \begin{split}
        \ket{\psi}  = e^{ \sum\limits_{n,m} \gamma_{nm}  {A}^{\dagger}_{n,H}  {A}^{\dagger}_{m,V}
        - H.c.}\ket{\text{vac}}\,.
    \end{split}
\end{align}
This a multi-mode squeezed state in the Gaussian-mode degree of freedom and a two-mode squeezed state in the polarization degree of freedom (indicated by the subscripts $H$ and $V$). To eliminate the polarization degree of freedom, we pass the state through a Mach-Zehnder interferometer with a HWP set to $45^{\circ}$ in one arm (see Appendix \ref{appA:Eliminating the Polarization Degree of Freedom for the Multi-mode State}). The output state is 
\begin{align}
    \begin{split}
        \ket{\psi^{\prime}}  ={}&\bigotimes_{j=1,2} e^{\frac{(-1)^j}{2}\sum\limits_{n,m}\gamma_{nm} A^{\dagger}_{n,j}A^{\dagger}_{m,j}
        - H.c.}\ket{\text{vac}} \,,
    \end{split}
\end{align}
which consists of two copies of a multi-mode squeezed state, all in the same polarization, where the subscript $j=1,2$ labels the output modes of the interferometer. Alternatively, on can pass the 

\subsection{Customizing the phase-matching function}\label{sec:cust}
We now turn to designing an appropriately shaped phase-matching function. Consider a nonlinear material whose nonlinearity can vary along the longitudinal direction $z$; this variation can be captured by a dimensionless function $g(z)$ (defined in Eq. \eqref{eq:scaled_g}). The function $g(z)$ can be transformed as follows:
\begin{equation}
\label{eq-theory-PMF definition}
   \Phi(\Delta k(\omega_s,\omega_i)) = \frac{1}{L} \int\limits_{-L/2}^{L/2} dz g(z)e^{i\Delta k(\omega_s,\omega_i) z}\,,
\end{equation}
which we can think of as an \emph{unnormalized} phase matching function. Here, $L$ is the length of the nonlinear material and $\Delta k(\omega_s,\omega_i) = k_p(\omega_s+\omega_i) - k_i(\omega_i) - k_s(\omega_s)$ is the \textit{phase mismatch}, where $k_J(\omega_J)=\omega_J n_J(\omega_J)/c$ and $n_J(\omega_J)$ is the refractive index for the mode $J$, $c$ is the speed of light and $J=s,i$. The unnormalized phase matching function $\Phi(\Delta k(\omega_s,\omega_i))$ is related to the normalized phase matching function $\phi(\omega_s-\omega_i)$ via Eq. \eqref{eq:phis}. When modelling custom-poled materials, it's easier to work with the unnormalized function $\Phi(\Delta k(\omega_s,\omega_i))$, and then to normalize the function numerically at the end. The target phase-matching function can then be written, in unnormalized form, as
\begin{align}\label{eq:target pmf}
    \Phi(\Delta k) =\sum_{m=-N}^{N} c_me^{-\frac{(\Delta k -\Delta k_0 - m\delta k)^2}{8\sigma_k^2}}\,,
\end{align}
where 
\begin{align}\label{eq:sigmak}
\sigma_k ={}& \frac{\sigma}{2}(k_s^\prime - k_i^\prime)\\
\delta k ={}&\label{eq:deltak} \frac{\delta\omega}{2}(k_s^\prime - k_i^\prime)\,,
\end{align}
and together with the Taylor expansion of $\Delta k(\omega_s,\omega_i)$ in Eq. \eqref{eq:app Dk} is equivalent to Eq. \eqref{eq:ch-5-general PMF}. 

To generate the desired phase matching function, we must determine the right form of $g(z)$. In principle, one can imagine varying $g(z)$ continuously. Such methods, however, don't exist for nonlinear crystals---in practice, for a given crystal, $g(z)$ is constrained to take on values of $\pm 1$ \cite{boyd2019nonlinear}. Experimentally, the sign of $g(z)$ can be alternated using a technique known as ferroelectric poling \cite{zukauskas2011fabrication,fejer1992quasi,imeshev2000ultrashort,arbore1997engineerable}, giving rise to individual domains.  In  $\Delta k$-space, each domain contributes a sinc function with a phase that depends on the domain position and orientation. By carefully arranging the positive and negative domains in $g(z)$, it is possible to interfere the sinc functions to construct phase-matching functions with almost arbitrary shapes. As with all quasi-phase-matching techniques, such as periodic poling, the resulting amplitude will necessarily be reduced when compared with intrinsically phase-matched materials. 

Several methods for designing appropriate domain configurations have been developed \cite{Dixon2013,chen2017efficient,chen2019indistinguishable,PhysRevApplied.12.034059,branczyk2011engineered,tambasco2016domain,dosseva2016shaping,graffitti2018independent,graffitti2017pure,chen2019indistinguishable}. Here, we focus on a variation of the algorithm originally proposed in \cite{tambasco2016domain} and further developed in \cite{graffitti2017pure, pickston2021optimised} (effects of  experimental imperfections in this approach were recently examined by Graffitti \emph{et al}. \cite{graffitti2018design}). In this approach, one computes an amplitude function---defined as the PMF, evaluated at a specific value of $\Delta k$, along the length of the crystal---then selects domains (one at a time from left to right) that bring the customized crystal's amplitude function closer to the target amplitude function. When the customized crystal's amplitude function closely approximates the target amplitude function at all points within the crystal, the customized crystal's PMF will also closely approximate the target PMF. This approach was recently demonstrated for an eight-peak PMF in KTP \cite{morrison2022frequency}.

In Appendix \ref{sec:design} we derive the following constraints on $c_m$:
\begin{equation}
    c_0 +2\sum_{m=1}^N c_m \le\sqrt{\frac{2}{\pi}} \frac{1}{L\sigma_k}.
\end{equation}
For the simple case where $c_m$ are all equal, the condition reduces to
\begin{equation}\label{eq:cneq}
    c_m = \sqrt{\frac{2}{\pi}} \frac{1}{L\sigma_kN_G},\hspace{2mm} -N\le m \le N\,,
\end{equation}
where $N_G$ is the number of Gaussian amplitudes.  It's desirable to maximize the  conversion efficiency, and thus to maximize $c_m$ within these constraints. 

Our implementation of the algorithm uses domain widths equal to the crystal's coherence length. As a result, the phase-matching function is constrained to be real and the coefficients should satisfy $c_m\approx c_{-m}$ (the approximation in this equality comes from a slight bias in the PMF discussed in Appendix \ref{sec:Bias in the PMF}). These restrictions would be lifted if a sub-coherence-length version of the algorithm was implemented \cite{graffitti2017pure}. In the next section, we demonstrate how to apply this technique to a specific example.  

\section{Example: 15-mode squeezed states} \label{sec:Example: 15-mode squeezed states}
\begin{figure}[b!]
    \centering
   \includegraphics[width=\linewidth]{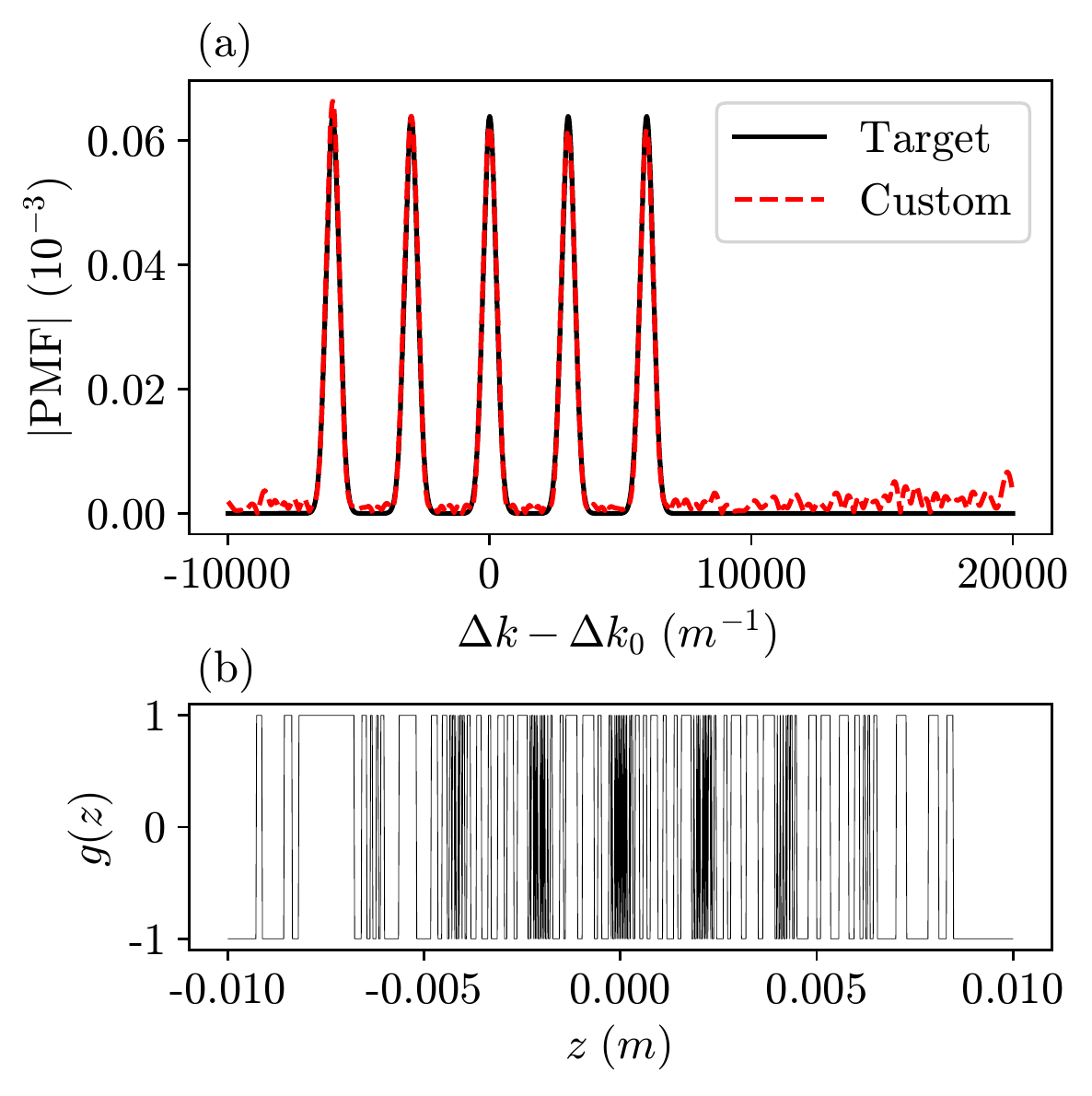}
    \caption{(a) Five-peak target phase-matching function as defined in Eq. \eqref{eq:target pmf} (with  $c_m = \sqrt{2}/(\sqrt{\pi}L\sigma_k5)$, $\sigma_k=2.5/L$ and $\delta k=24\sigma_k$) and corresponding custom phase-matching function generated by the custom-poled crystal shown in (b). We took $\sigma_k=2.5/L$ to ensure that the target nonlinearity profile fits within the length of the crystal. The custom-poled crystal has $N=1073$ domains of width $w=18.63\mu m$. The domain width was chosen to match the phase-matching conditions of Type-II KTP in the symmetric group-velocity-matched regime. }
    \label{fig:1}
\end{figure}

As an example, we customize a 2cm Potassium titanyl phosphate (KTP) crystal to generate a PMF with five Gaussian peaks centered at $\Delta k_0-m\delta k$,  where $m=(0,\pm 1,\pm 2)$ and $\delta k = 24\sigma_k$. We set $\sigma_k= 2.5/L$ to ensure that the target nonlinearity profile fits within the length
of the crystal. The target PMF is given by Eq. \eqref{eq:target pmf} with $N=2$. Figure \ref{fig:1} shows the generated PMF compared to the target PMF, as well as the resulting domain configuration $g(z)$. Notice the expected slight bias in the generated PMF discussed in Appendix \ref{sec:Bias in the PMF}.

\begin{figure}[h!]
    \centering
    \includegraphics[width=\columnwidth]{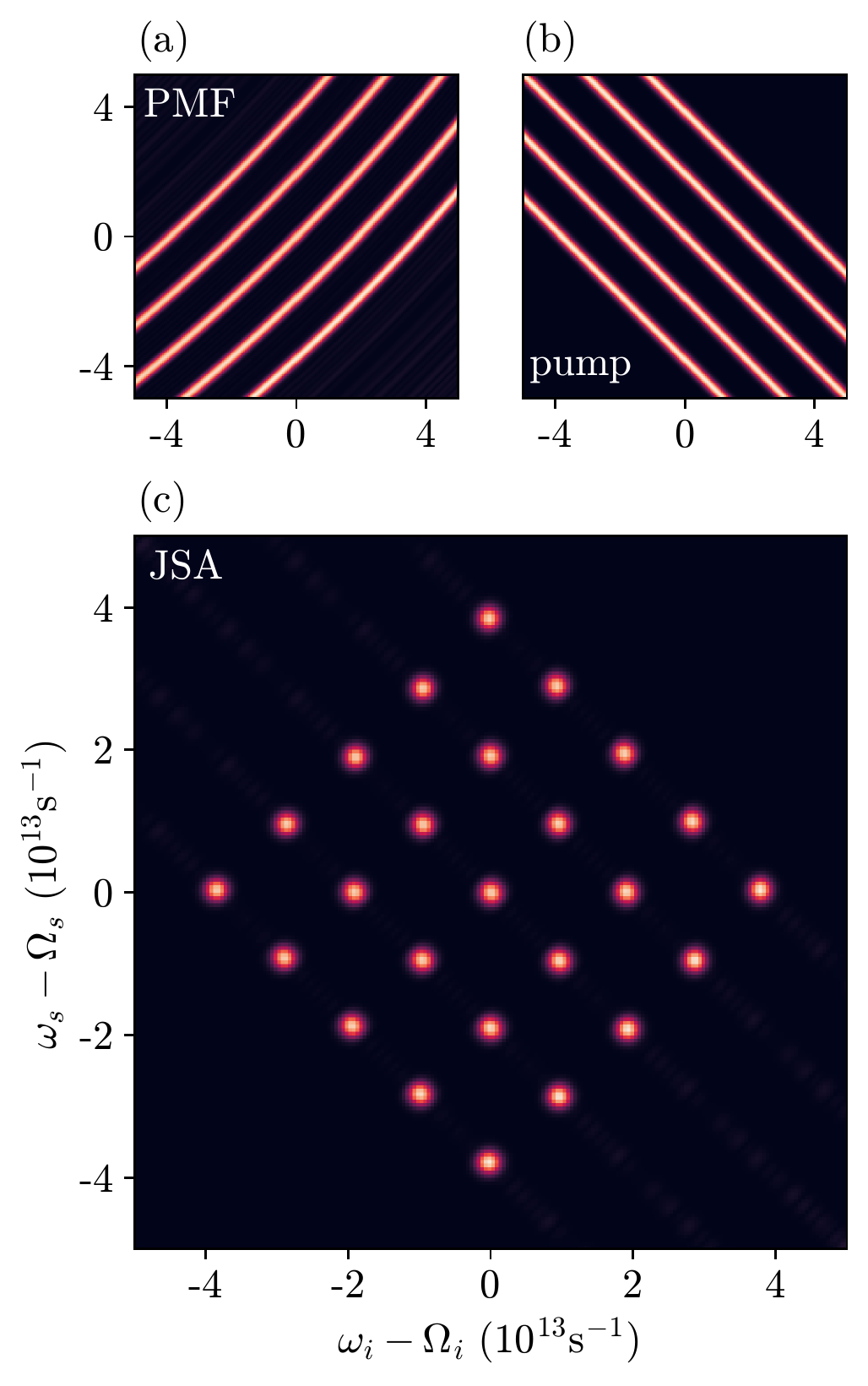}
    \caption{(a) Five-peak phasematching function (PMF) as defined by Eq. \eqref{eq:ch-5-general PMF} with $\sigma$ and $\delta\omega$ related to $\sigma_k$ and $\delta k$ in Fig \ref{fig:1} via Eqs. \eqref{eq:sigmak} and \eqref{eq:deltak}. (b) Five-peak pump function  as defined in Eq. \eqref{eq-generalization-general PUMP}  with the same $\sigma$ and $\delta\omega$ as the PMF. (c) The resulting joint spectral amplitude (JSA) as defined in Eq. \eqref{eq:JSAreal}. }
    \label{fig:2}
\end{figure}

We design a pump function with five peaks centred at 
$ \omega = \Omega_{p}-n\delta\omega$,  where $n=(0,\pm 1,\pm 2)$. As with the PMF, we take all coefficients $c_m$ to be real, equal and given by the restriction in Eq. \eqref{eq:cneq}. We take $\sigma=2\sigma_k/(k_p'-k_s')$ and to  ensure that the spacing is the same as for the PMF, we take $\delta \omega = 24\sigma$. Then using Eq. \eqref{eq:sigmak} and $\sigma_k= 2.5/L$, the pump bandwidth in frequency is $\sigma/2\pi = 0.127\text{THz}$ and in time is on the order of 7ps. The resulting JSA with corresponding PMF and pump is shown in Figure \ref{fig:2}. Each peak in the JSA corresponds to a term in Eq. \eqref{eq:polstate}. Notice that there is a slight bend in the PMF due to dispersion (for these plots, we used the full Sellmeier equations rather than the first-order approximation of $\Delta k$). Too much dispersion will reduce the effectiveness of this technique, but in the example shown here, the effect on the JSA is negligible. 

It's possible to tune the amplitudes $\gamma_{nm}$ to some extent by tuning the height of the peaks in the pump function (the height of the peaks in the PMF are fixed for a given crystal). Tuning the peak of the pump function is equivalent to scaling the height of the modes that lie along the same anti-diagonal in the JSA. 

The JSA in Fig \ref{fig:2} has 25 peaks, but it corresponds to a 15-mode squeezed state because the PMF is symmetric in its amplitudes. The JSA is also symmetric along the line $\omega_s = \omega_i$. There are 10 amplitudes above the line $\omega_s = \omega_i$ which are all centered at different frequencies and thus correspond to two-mode squeezing terms. Each amplitude along the diagonal is centered at the same center frequency and thus corresponds to each of the 5 single-mode squeezing terms. 

\section{Extension: Tunable hybrid squeezed states} \label{sec:Tunable hybrid squeezed states}
In this section, we introduce, and describe the generation of hybrid squeezed states, which have features of single-mode and two-mode squeezed vacuum states. 
Consider the special case of multi-mode squeezed state (Eq. \eqref{eq:mmsv v1}) restricted to two modes:
\begin{equation}
    \ket{\textsc{thss}} = e^{\frac{\beta_{11}}{2}a^{\dagger}_1a^{\dagger}_1 + \beta_{12} a^{\dagger}_1a^{\dagger}_2 +   \frac{\beta_{22}}{2}a^{\dagger}_2a^{\dagger}_2 - \hc}\ket{\text{vac}}\,,
    \label{eq:ch-4-STMSV state intro to ch 4}
\end{equation}
where we restricted $\beta_{12} = \beta_{21}$. We can tune the constants to continuously move between a single-mode squeezed state ($\beta_{12}=\beta_{22}=0$), a two-mode squeezed state ($\beta_{11}=\beta_{22}=0$) or a product of two single-mode squeezed states ($\beta_{12}=0$). We call this a tuneable hybrid squeezed state (THSS).

\begin{figure}[h]
    \centering
        \includegraphics[width=\columnwidth]{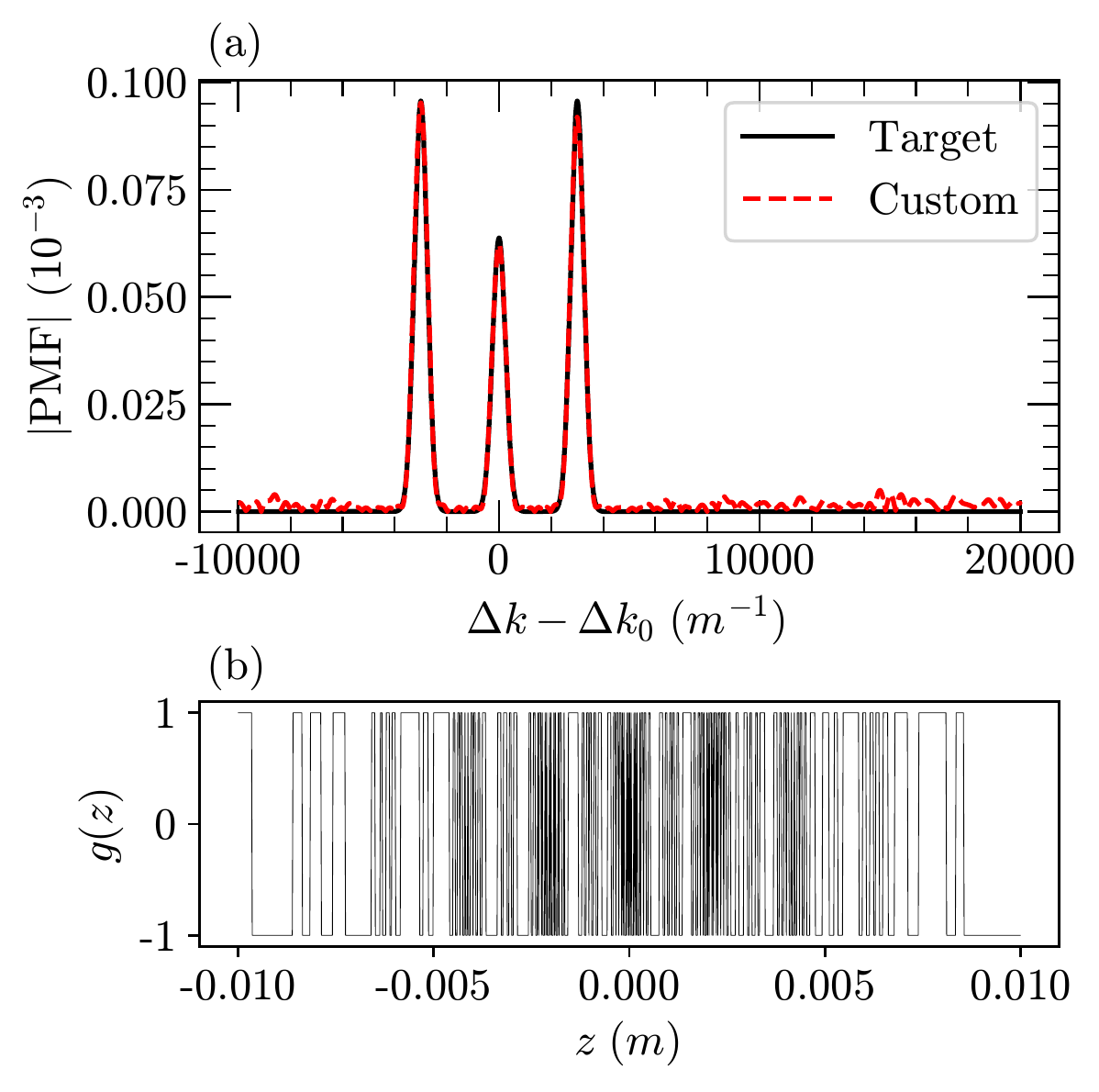}
    \includegraphics[width=\columnwidth]{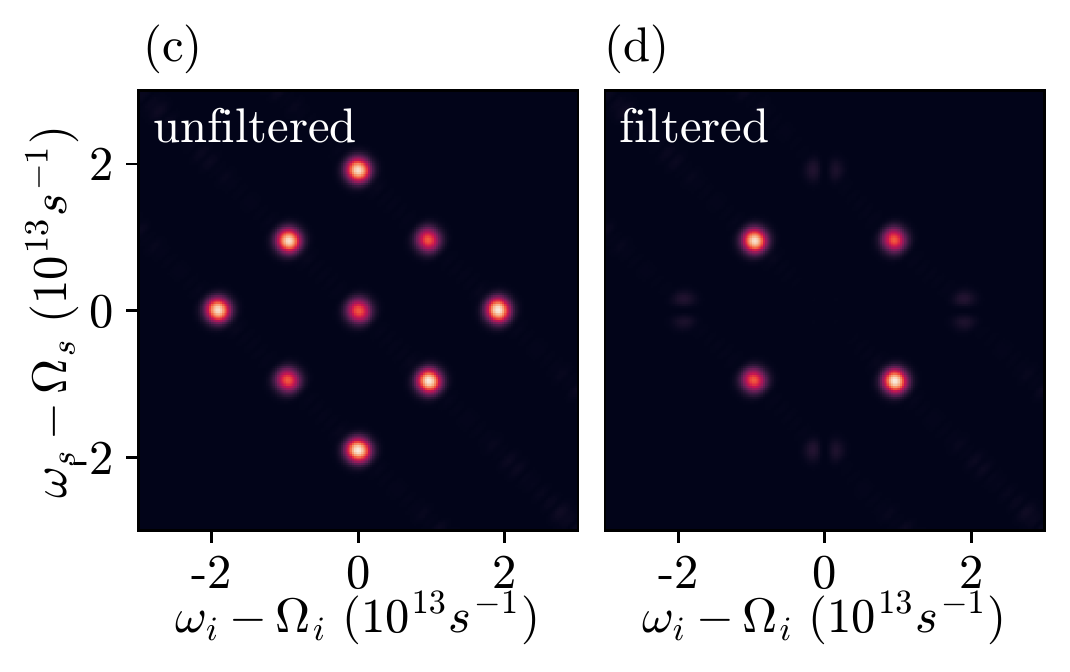}
    \caption{(a) Three-peak target phase-matching function as defined in Eq. \eqref{eq:target pmf} (with
    $c_0= \sqrt{2}/(\sqrt{\pi}L\sigma_k
    5)$, $c_1=c_{-1} = 3c_0/2$) and corresponding custom phase-matching function generated by the custom-poled crystal shown in (b). Parameters $\sigma_k$, $\delta k$, $N$ and $w$ are the same as in Figure \ref{fig:1}. (c) The resulting joint spectral amplitude (JSA) as defined in Eq. \eqref{eq:JSAreal} for a 3-peak phasematching function (PMF) and a 3-peak pump function with $\sigma$ and $\delta\omega$ the same as in Figure (d) The JSA in (c) with a filter applied to both the signal and idler modes, with a transmission function $(1-\exp(-(\omega_s-\Omega_s)^2/4\sigma_f^2))\times(1-\exp(-(\omega_i-\Omega_i)^2/4\sigma_f^2))$ where $\sigma_f = 2\sigma$.}
    \label{fig:3}
\end{figure}

Using the pulse-mode encoding introduced in Section \ref{sec:Pulsed-mode encoding}, a JSA corresponding to a THSS has four amplitudes located at, say,  $(\Omega_{-1},\Omega_{-1})$, $(\Omega_{-1},\Omega_{1})$, $(\Omega_{1},\Omega_{-1})$ and $(\Omega_{1},\Omega_{1})$, each corresponding to a squeezing term in Eq. \eqref{eq:ch-4-STMSV state intro to ch 4}.

To generate such a JSA, we create a three-peak PMF and a three-peak pump function, to produce a 9-peak JSA show in figure \ref{fig:3}. The JSA in Figure \ref{fig:3} has extra squeezing terms located at $\Omega_0$. However, the full state generated from the JSA in Figure \ref{fig:3} can be written as a product of two squeezing operators acting on the vacuum state given by
\begin{align}\nonumber
    \ket{\psi} ={}&\left(e^{\frac{\gamma_{11}}{2}A_{1}^\dagger A_{1}^\dagger  +\frac{\gamma_{-11}}{2}A_{{-}1}^\dagger A_1^\dagger+ \frac{\gamma_{{-}1{-}1}}{2}A_{{-}1}^\dagger A_{{-}1}^\dagger -\hc }\right)\\\label{eq:sq22}
    \times{}&\left(e^{\gamma_{{-}20}A_{{-}2}^\dagger A_0^\dagger  +\frac{\gamma_{00}}{2}A_0^\dagger A_0^\dagger +\gamma_{20}A_{2}A_{0}-\hc }\right)\ket{\text{vac}}\,.
\end{align}
We can always decompose the state in this way because the modes are orthogonal and thus $[A_n,A_m^\dagger] = \delta_
{nm}$. This is crucial because it means the state corresponding to Figure \ref{fig:3} can be written as $\ket{\psi} = \ket{\textsc{thss}}\otimes\ket{\phi}$ and thus $\ket{\phi}$ can be traced out without degrading the purity of the desired modes. To trace out the state $\ket{\phi}$ in practice, we can pass the 9-peak JSA state through a spectral filter that blocks frequencies around $\Omega_0$. The resulting JSA is shown in Figure \ref{fig:3} d). 

This hybrid squeezed state  can be tuned in real time by tuning the amplitude of each term in the pump function. This is because each term in the pump function uniquely corresponds to terms in the squeezing operator; specifically, each peak centred at $(\Omega_n,\Omega_m)$ corresponds to the squeezing term associated with the $A_nA_m$ operators in \eqref{eq:sq22}. The middle peak in the pump amplitude generates two JSA amplitudes at $(\Omega_{-1},\Omega_{1})$ and $(\Omega_1,\Omega_{-1})$ but these both contribute to the same amplitude in the squeezed state. Since the pump is a function of three centre frequencies with no overlap we can independently vary the amplitude of all three and thus tune the squeezing parameters of the final squeezed state independently.  
\section{Conclusion}\label{sec:Conclusion}
We proposed a method for generating multi-mode squeezed states of light encoded in Gaussian-like frequency modes. This method differs from related methods \cite{pfister2004multipartite, menicucci2007ultracompact, alexander2016one,zhu2021hypercubic,Yeoman1993TwomodeSG,van2000multipartite,ansari2018tailoring,fabre2020generation} in several ways that might make it advantageous, depending on the application.

The proposed method relies on customizing the joint spectral properties of light generated via spontaneous parametric down conversion, which requires two independent ingredients. The first is the spectral engineering of light incident on the crystal, which we took as a given. The second is the engineering of the nonlinear crystal to have desired phasematching properties (captured by the \emph{phasematching function}), for which we used an algorithm \cite{graffitti2017pure} that takes as an input, the target phasematching function, and outputs a binary string that defines poles in a ferroelectrically-poled crystal \cite{yamada1993first}.

We used this method to design two kinds of squeezed states. The first is an $N$-mode squeezed state, which could generalize various quantum information protocols \cite{gessner2018sensitivity,sokolov2004squeezed,lian2007continuous}. The second is a tuneable hybrid squeezed state, which could make good resource states for the generation of non-Gaussian states via post-selection, particularly in situations where mode tunability is desired, e.g., to compensate for loss.

Future work could investigate the effects of time-ordering on the JSA \cite{quesada2017you}, suppression of modes within the crystal using photonic stop bands \cite{helt2017parasitic}, and possible applications in other areas of research, such as, multi-parameter quantum metrology and multi-channel quantum imaging. We expect that exploring new quantum states of light, encoded in new degrees of freedom, such as those proposed here will be a fruitful path toward discovering new quantum applications. 

The Jupyter notebook used to generate the results in this paper can be found at \url{https://github.com/abranczyk/custom-poling}.

\textit{Acknowledgments} --- AMB thanks Austin Lund, Olivier Pfister, and Alessandro Fedrizzi for valuable discussions. Research at Perimeter Institute is supported in part by the Government of Canada through the Department of Innovation, Science and Economic Development Canada and by the Province of Ontario through the Ministry of Colleges and Universities. We acknowledge the support of the Natural Sciences and Engineering Research Council of Canada (RGPIN-2016-04135). 

\bibliographystyle{unsrt}
\bibliography{references}
\newpage
\appendix
\input{Appendices}

\section{Review of Squeezed states}\label{sec:Background}
Here we briefly review single-mode, two-mode and multi-mode squeezed states. A squeezed state prepared in a single mode $a_0$, can be written as \cite{scully_zubairy_1997}
\begin{equation}
\begin{split}
    \label{eq:smsv}
    \ket{\textsc{smsv}} &= e^{\frac{1}{2}\beta_{00}  {a}^\dagger_0  {a}^\dagger_0 -\hc}\ket{\text{vac}}\,,
\end{split}
\end{equation}
where $\beta_{00}$ is the complex single mode squeezing parameter. If we define the uncertainty of an operator $A$ by $(\Delta A)^2 = \bra{\textsc{smsv}}(A - \ex{A})^2\ket{\textsc{smsv}}$, the single-mode squeezed state has the property that it minimizes the uncertainty relation below the quantum noise limit. This is taken advantage of in quantum sensing experiments, where one decreases the phase uncertainty to measure changes in distance beyond the quantum noise limit \cite{lawrie2019quantum,aasi2013enhanced}. 

A two-mode squeezed state prepared in modes $a_0$ and $a_1$, can be written as  \cite{scully_zubairy_1997}
\begin{equation}
    \begin{split}
    \label{eq:tmsv}
        \ket{\textsc{tmsv}} &= e^{\beta_{01}
        {a}^\dagger_0  {a}^\dagger_1 -\hc}\ket{\text{vac}}\,,
    \end{split}
\end{equation}
where $\beta_{01}$ is the complex two-mode squeezing parameter. If one considers the effect that the squeezing operator has on the variance of the sum and difference of each modes conjugate variables, one finds a similar squeezing effect as for the single-mode state. For the two operators $A$ and $B$, the two-mode squeezed state minimizes the uncertainty relation between the conjugate variables $A-B$ and $A+B$ below the quantum noise limit. The reduction in noise between $A-B$ generates a high degree of correlation, and in the limit when $|\beta_{01}|\to \infty$, two-mode squeezed states are the continuous variable extension of maximally entangled Bell states \cite{wang2007quantum}. The entanglement properties of two-mode squeezed state can be used for various quantum information protocols, such as, quantum teleportation and quantum cryptography \cite{milburn1999quantum,hillery2000quantum}.

The squeezed states above can be generalized to $N$-partite entangled states as follows:
\begin{align}
\label{eq:mmsv v1}
    \ket{\textsc{mmsv}} = e^{\frac{1}{2}\sum\limits_{n,m}\beta_{nm}{a}_{n}^{\dagger}{a}^{\dagger}_{m} -\hc}\ket{\text{vac}},
\end{align}
where the double sum in the exponential ranges  from $n,m =-N,-N+1,...,0,...,N-1,N$ and due to symmetry we take $\beta_{nm} = \beta_{mn}$. These \emph{multi-mode squeezed states} can be used to generalize various quantum information protocols \cite{armstrong2012programmable,epping2017multi}.

\section{Spontaneous Parametric Downconversion}\label{sec:SPDC}

In this section we review the nonlinear-optical process known as spontaneous parametric downconversion (SPDC). In a SPDC process the input photons, typically called pump photons, are each ``downconverted'' into two daughter photons, called the signal and idler. The signal and idler photons satisfy energy and momentum conservation with the pump photon given by
\begin{align}
    \omega_p = \omega_s + \omega_i, \hspace{5mm}k_p(\omega_p) = k_s(\omega_s) + k_i(\omega_i)
    \label{eq-theory-SPDC energy and momentum conservation}
\end{align}
where $\omega_J$ and $k_J(\omega_J)$ are the frequency and wave vectors. The three fields of interest are labeled by $J = p,s,i$, denoting the pump, signal and idler respectively. A detailed derivation of SPDC was given in \cite{yang2008spontaneous} which we follow. 

For a Type II downconversion process---in the rotating wave approximation, assuming the material is in an effective 1D structure where the field doesn't vary in the orthogonal direction of area $A$, with a non-linear coefficient  $\chi^{(2)}(z)$ that varies along the longitudinal direction $z$---the nonlinear Hamiltonian in the interaction picture is
\begin{align}
    \begin{split}
          H_{I}(t) =& -\hbar \int_0^{\infty} d\omega_p d\omega_i d\omega_s  {c}_{V}(\omega_p)  {a}^{\dagger}_{V}(\omega_i)  {a}_{H}^{\dagger}(\omega_s)e^{i(\omega_s + \omega_i - \omega_p)t} 2L\sqrt{\frac{ \hbar\omega_p \omega_i \omega_s }{(4\pi)^3\epsilon_0 A c^3 n_p(\omega_p) n_s(\omega_s) n_i(\omega_i)}}  \\
        &\hspace{50mm}\times\frac{1}{L}\int_{-L/2}^{L/2} dz \chi^{(2)}(z) e^{i(k_p(\omega_p) - k_i(\omega_i) - k_s(\omega_s))z} + H.c..
        \end{split}
\end{align}
For nonlinear materials we will be considering, $\chi^{(2)}(z)$ is constant over a specified domain length and given by $\chi^{(2)}(z) = \pm\chi^{(2)}$. We find it useful to define
\begin{equation}\label{eq:scaled_g}
    g(z) = \frac{\chi^{(2)}(z)}{\chi^{(2)}}
\end{equation}
to be the scaled nonlinearity function which has two values given by $g(z) = \pm1$. If we take the initial ket $\ket{\psi(-\infty)}$ to be a coherent state in the pump mode and the signal and idler vacuum then in the undepleted pump approximation (where we assume the pump light is unchanged with the removal of a photon) we can make the substitution that 
\begin{align}
      c_V(\omega_p) \to\sqrt{|\alpha_p|}\alpha(\omega_p),
\end{align}
where $\alpha(\omega_p)$ is the normalized frequency distribution of the pump laser, which satisfies
\begin{equation}
    \label{eq:norm pump}
    \int |\alpha(\omega)|^2d\omega = 1, 
\end{equation}
and $|\alpha_p|^2$ is the number of photons in the pump. Then the Hamiltonian is given by
\begin{align}
    \begin{split}
          H_{I}(t) =& -\hbar \int_0^{\infty} d\omega_p d\omega_i d\omega_s  \mathcal{A}(\omega_i,\omega_s,\omega_p)\alpha(\omega_p)\Phi(\omega_s,\omega_i,\omega_p)  {a}^{\dagger}_{V}(\omega_i)  {a}_{H}^{\dagger}(\omega_s)e^{i(\omega_s + \omega_i - \omega_p)t} + H.c.,
        \end{split}
\end{align}
where we set
\begin{equation}\label{eq:A}
    \mathcal{A}(\omega_s,\omega_i,\omega_p) = 2L\chi^{(2)}\sqrt{\frac{ \sqrt{|\alpha_p|}\hbar\omega_p \omega_i \omega_s }{(4\pi)^3\epsilon_0 A c^3 n_p(\omega_p) n_s(\omega_s) n_i(\omega_i)}},
\end{equation}
and define
\begin{equation}
    \label{eq:unnormalized PMF}
    \Phi(\Delta k(\omega_i,\omega_s,\omega_p)) = \frac{1}{L}\int\limits_{-L/2}^{L/2} dz g(z) e^{i\Delta k(\omega_p,\omega_s,\omega_i)z},
\end{equation}
to be the phasematching function (PMF) and set $\Delta k(\omega_p,\omega_s,\omega_i) = k_p(\omega_p) - k_i(\omega_i) - k_s(\omega_s)$ to be the \textit{phase mismatch}.

In the interaction picture the states evolves according to interaction Hamiltonian given by \cite{shankar2012principles}
\begin{align}
    i\hbar \frac{d\ket{\psi(t)} }{dt} =   H_I(t)\ket{\psi(t)},
\end{align}
which has a formal solution of 
\begin{align}
    \ket{\psi(\infty)} = \mathcal{T}[e^{\frac{-i}{\hbar}\int\limits_{-\infty}^{\infty} dt   H_I(t)}] \ket{\psi(-\infty)},
\end{align}
where $\mathcal{T}$ is the time-ordering operator and $\ket{\psi(\infty)}$ is the final state. Since the interaction Hamiltonian does not commute with itself at different times, we cannot in general drop the time ordering operator. It was shown in \cite{quesada2017you}, that the time ordering leads to non trivial effects but only in the high pump power regime. For this work, we will assume low pump powers and not worry about these time ordering effects, i.e. we drop the time-ordering operator. Then integrating the
interaction Hamiltonian with respect to $t$ introduces an energy conserving delta function $\delta(\omega_s + \omega_i - \omega_p)$ which we use to evaluate the $\omega_p$ integral. Then the state is given by 
\begin{equation}
    \ket{\psi} = e^{i\int d\omega_sd\omega_i\mathcal{A}(\omega_i,\omega_s)\alpha(\omega_i + \omega_s)\Phi(\Delta k(\omega_i,\omega_s))a_H^\dagger(\omega_i)  a_V^\dagger(\omega_s)+ H.c.}\ket{\text{vac}},
\end{equation}
where we made the simplification $\mathcal{A}(\omega_s,\omega_i,\omega_i + \omega_s)\to\mathcal{A}(\omega_s,\omega_i)$ and $\Phi(\Delta k(\omega_i,\omega_s,\omega_i + \omega_s))\to\Phi(\Delta k(\omega_i,\omega_s))$ with
\begin{equation}
    \Delta k(\omega_i,\omega_s) = k_p(\omega_i + \omega_s) - k_i(\omega_i) - k_s(\omega_s),
\end{equation}
being the \emph{phase mismatch} and $k_J(\omega_J) = \omega_Jn_J(\omega_J)/c$. Lastly, for the frequency range of interest, to good approximation we can evaluate the frequency dependent term $\mathcal{A}(\omega_s,\omega_i)\approx\mathcal{A}(\Omega_s,\Omega_i)=\mathcal{A}$, where $\Omega_s$ and $\Omega_i$ are the center frequencies of the signal and idler.

To push the equations further, we expand the phase mismatch $\Delta k$ to first order around the center frequencies $\Omega_s,\Omega_i, \Omega_p = \Omega_s + \Omega_i$  such that 
\begin{equation}
    \Delta k(\omega_i,\omega_s)=\Delta k_0+k_p'(\omega_p-\Omega_p) - k_i'(\omega_i-\Omega_i) - k_s'(\omega_s-\Omega_s)
\end{equation}
where we set $\Delta k_0=k_p(\Omega_s+\Omega_i) - k_i(\Omega_i) - k_s(\Omega_s)$ and the first-order derivatives are $k_j'=\partial{k_j(\omega_j)}/\partial{\omega_j}|_{\omega_j=\Omega_j}$, and energy conservation ensures that $\omega_p=\omega_i+\omega_s$ and $\Omega_p=\Omega_i+\Omega_s$. Next, we work in the symmetric group-velocity matching regime, such that  $k_p'=(k_s'+k_i')/2$, then with these choices, we are left with 
\begin{align}
    \label{eq:app Dk}
    \Delta k(\omega_i,\omega_s) ={}&\Delta k_0+\frac{k_s'-k_i'}{2}((\omega_i-\Omega_i)-( \omega_s - \Omega_s))\,.
\end{align}
We note that $\Delta k(\omega_i,\omega_s)$ is now a function of the variable $\omega_i-\omega_s$ so we take $\Delta k(\omega_i,\omega_s)\to \Delta k(\omega_i-\omega_s)$ and therefore $\Phi(\Delta k(\omega_i,\omega_s))\to\Phi(\omega_i-\omega_s)$ which for typical phase matching functions will be peaked along the diagonal. Calculating the square integral of $\Phi(\omega)$ we find it is not yet normalized and set its value to be $\mathcal{N}^2$ for some $\mathcal{N}$, therefore
\begin{equation}
    \int d\omega |\Phi(\omega)|^2 = \mathcal{N}^2.
\end{equation}
To normalize $\Phi(\omega)$ we define 
\begin{equation}\label{eq:phis}
    \phi(\omega) = \frac{2}{\mathcal{N}}\Phi(\omega),
\end{equation}
then the squeezed state is given by
\begin{equation}
    \ket{\psi} = e^{i\gamma\int d\omega_sd\omega_i f(\omega_s,\omega_i)a_H^\dagger(\omega_i)  a_V^\dagger(\omega_s)+ H.c.}\ket{\text{vac}},
\end{equation}
where we defined the squeezing parameter $\gamma$ to be $\gamma = \mathcal{A}\mathcal{N}$ and defined the join spectral amplitude (JSA) as
\begin{align}
    f (\omega_s,\omega_i) =\alpha(\omega_i + \omega_s)\phi(\omega_i-\omega_s),
\end{align}
which due to the normalization in Eq. \eqref{eq:norm pump} and  \eqref{eq:phis} satisfies the normalization
\begin{equation}
    \int d\omega_sd\omega_i|f(\omega_s,\omega_i)|^2=1.
\end{equation}

\section{Eliminating the Polarization Degree of Freedom for the Multi-mode State}
\label{appA:Eliminating the Polarization Degree of Freedom for the Multi-mode State}

In Eq. \eqref{eq:JSAreal2} we decomposed the JSA into a set of frequency modes with a frequency distribution given by $G_n(\omega)$. Although we have provided a decomposition there is still the polarization degree of freedom which we need to remove. To eliminate the polarization degree of freedom we apply the following set of transformation. We begin by applying a polarization beam splitter which has the transformation property 
\begin{equation}
    \begin{split}
        &  {U}_{PBS}^{\dagger}  {a}_H^{(1)}(\omega)   {U}_{PBS}  =   {a}_{H}^{(1^{\prime})}(\omega) ,\\
        & {U}_{PBS}^{\dagger}  {a}_V^{(1)}(\omega)   {U}_{PBS} =  {a}_{V}^{(2^{\prime})}(\omega), \\
        & {U}_{PBS}^{\dagger}  {a}_H^{(2)}(\omega)   {U}_{PBS} =  {a}_{H}^{(1^{\prime})}(\omega) ,\\
        & {U}_{PBS}^{\dagger}  {a}_V^{(2)}(\omega)   {U}_{PBS} =  {a}_{V}^{(2^{\prime})}(\omega) ,\\
    \end{split}
\end{equation}
where the superscripts denote the input and output spatial modes. Next we apply
a half-wave-plate to one of the spatial modes set to $45^{\circ}$ using the following transformation property
\begin{equation}
    \begin{split}
        U_{HWP}^{\dagger}(45^{\circ}) {a}_H (\omega)U_{HWP}(45^{\circ}) &= i{a}_V (\omega)\\
        U_{HWP}^{\dagger}(45^{\circ})  {a}_V(\omega) U_{HWP}(45^{\circ}) &= i{a}_H (\omega).
    \end{split}
\end{equation}
Finally, we recombine the modes using a 50:50 beam splitter set to $0^{\circ}$ with the transformation given by
\begin{equation}
    \begin{split}
       U_{BS}^{\dagger}(0^{\circ})  {a}^{(1)}(\omega) U_{BS}(0^{\circ}) & = \frac{1}{\sqrt{2}}( {a}^{(1)}(\omega) +  {a}^{(2)}(\omega))\\
        U_{BS}^{\dagger}(0^{\circ})  {a}^{(2)}(\omega) U_{BS}(0^{\circ})& = \frac{1}{\sqrt{2}}(-  {a}^{(1)}(\omega) +  {a}^{(2)}(\omega)).
    \end{split}
    \label{eq-theory-BS transformation}
\end{equation}

After applying these three transformations and using the property that $\gamma_{nm} = \gamma_{mn}$ and dropping the polarization degree of freedom since they are all the same the final state is given by
\begin{align}
    \begin{split}
        \ket{\psi^{\prime}}  =& ^{ \frac{1}{2}\sum\limits_{n,m} \gamma_{nm} A^{\dagger,(1)}_{n}A^{\dagger,(1)}_{m}
        - H.c.}\ket{\text{vac}}  \otimes e^{ -\frac{1}{2}\sum\limits_{n,m} \gamma_{nm} A^{\dagger,(2)}_{n}A^{\dagger,(2)}_{m}
        - H.c.}\ket{\text{vac}},
    \end{split}
\end{align}
which is two copies of a multi-mode squeezed state as defined in Eq. \eqref{eq:mmsv An} in two different spatial degrees of freedom.

\section{Bias in the PMF}
\label{sec:Bias in the PMF}
For a crystal of length $L$ the approximate phasematching function is given by
\begin{equation}
    \Phi_a(\Delta k) =\frac{1}{L} \int_{-L/2}^{L/2} dz g_a(z)e^{i \Delta k z},
\end{equation}
where $g_a(z)$ is the approximate nonlinearity function specified by the poling algorithm and is either $\pm1$. The  phasematching function can be written as a coherent sum by expanding the integral over $z$ into each domain by
\begin{equation}
    \Phi_a(\Delta k) = \frac{1}{L}  \sum_{n=0}^{n=N} g_n \int\limits_{-\frac{L}{2}+n l_c}^{-\frac{L}{2}+(n+1)l_c}e^{i\Delta k z}dz,
\end{equation}
where $l_c$ is the coherence length, $Nl_c = L$, and $g_n \equiv g_a (-\frac{L}{2}+n l_c \le z \le -\frac{L}{2}+(n+1)l_c )$ is the nonlinearity within each domain. Then evaluating the integral and simplifying the approximate phasematching function is given by
\begin{equation}
    \Phi_a(\Delta k) = \frac{l_c}{L}\text{sinc}\left(\frac{\Delta k l_c}{2}\right)e^{i\Delta k l_c/2}e^{-i\Delta kL/2}\sum_{n=0}^{n=N} g_n e^{i\Delta k l_c n}.
\end{equation}
Since the algorithm we are using to determine $g_n$ is only for real phasematching functions we know the imaginary part will sum to zero and the nonzero contribution is only from the real part which is given by
\begin{equation}
    \label{eq:real part-approx. pmf}
    \Re{\Phi_a(\Delta k)} = \frac{l_c}{L}\frac{\text{sin}\left(\frac{\Delta k l_c}{2}\right)}{\left(\frac{\Delta k l_c}{2}\right)}\sum_{n=0}^{n=N} g_n \cos\left(\frac{\Delta k l_c}{2}(2n+1-N)\right),
\end{equation}
which is manifestly symmetric about $\Delta k = 0$. However, the phasematching function we are designing has peaks at $\Delta k = \Delta k_0 \pm m\delta k$ which is centered at $\Delta k_0$ ensuring the nonlinear generation is phasematched. Evaluating the approximate phasematching function in this vicinity leads to a $1/(\Delta k_0 \pm m\delta k)$ dependence which is no longer symmetric and leads to a bias on the left and right sides of $\Delta k_0$ shown in Figure \ref{fig:1}. 

Although the phasematching function bias for large or smaller values of $\Delta k$ leads to less accurate PMFs; by decreasing the coherence length the affect of the bias is lessened, and the PMF is better approximated. Why can understand why this necessarily follows in two ways: As we decrease the coherence length the sinc function prefactor in Eq. \eqref{eq:real part-approx. pmf} becomes more broad as is therefore constant for larger values of $\Delta k$; minimizing the bias. Secondly, by decreasing the coherence length we are increasing the ``resolution'' of our tracking algorithm which increases the accuracy of the approximated PMF. To Further increase the accuracy of the generated PMFs we can move to sub-coherence domain engineering which was discussed in detail by Graffitti \emph{et al.} in \cite{graffitti2017pure}.

\section{Design Considerations}
\label{sec:design}

Quasi-phase-matched periodically-poled crystals are known to generate amplitudes  reduced by a factor of $2/\pi$ when compared with their phase-matched counterparts \cite{tambasco2016domain}. For example, if the unnormalized phase-matching function for a phase-matched crystal of length $L$ is $\Phi(\Delta k)=\frac{1}{L}\int_{-L/2}^{L/2}e^{-i\Delta k z}dz=\mathrm{sinc}(\Delta k L/2)$, then for a periodically-poled crystal, it is $\Phi(\Delta k)\approx (2/\pi)\mathrm{sinc}(\Delta k L/2)$. This is due to the interference behind the quasi-phase-matching effect. The phase-matching function is at its maximum value of $(2/\pi)$ at the phase-matching condition ($\Delta k=0$). The amplitude along the longitudonal direction $z$ is $(2z/L\pi)$, which defines the maximum slope for the amplitude inside the crystal. 

We can use the maximum slope restriction to put bounds on the coefficients $c_n$ that scale each Gaussian peak in the target PMFs by making sure that the gradient of the amplitude function satisfies:
\begin{equation}
    \frac{d}{dz}A_t(z,\Delta k_0) \le  \frac{d}{dz}\left(\frac{2z}{\pi L}\right)= \frac{2}{\pi L}.
\end{equation}
The amplitude of the PMF for a given $\Delta k_0$ throughout the crystal is given by 
\begin{equation}\label{eq:AAAA}
    A_t(z,\Delta k_0) = \frac{1}{L}\int\limits_{-L/2}^{z}dz^\prime g_t(z^\prime)e^{i\Delta k_0z^\prime},
\end{equation}
then the target nonlinearity function $g_t(z)$ can be found by inverting Eq. \eqref{eq-theory-PMF definition}:
\begin{equation}\label{eq:gggg}
    g_t(z) = \frac{L}{2\pi}\int\limits_{-\infty}^\infty d(\Delta k)\Phi_t(\Delta k)e^{-i\Delta k z}\,.
\end{equation}
We now pick a specific form for the target PMF $\Phi_t(\Delta k)$ with $N_G = 2N+1$ Gaussian amplitudes,
\begin{equation}
\label{eq:target PMF many amplitudes}
    \Phi_t(\Delta k)  =\sum_{m=-N}^{N}c_{m}e^{-\frac{(\Delta k -\Delta k_0 - m\delta k)^2}{8\sigma^2_k}}\,,
\end{equation}
and insert it into  \eqref{eq:gggg} to give
\begin{equation}\label{eq:phit}
\begin{split}
    g_t(z) \approx &\frac{2L\sigma_k}{\sqrt{2\pi}}e^{-2z^2\sigma_k^2}e^{-i\Delta k_0 z}\left(c_0 +2\sum_{m=1}^Nc_m\cos(m\delta k z)\right),
\end{split}
\end{equation}
where we used the approximation that the amplitudes of the PMF are symmetric ($c_m=c_{-m}$). Now inserting \eqref{eq:phit} into \eqref{eq:AAAA} and taking the derivative with respect to $z$ we find that the coefficients must satisfy
\begin{equation}
    \frac{2\sigma_k}{\sqrt{2\pi}}e^{-2z^2\sigma_k^2/2}\left(c_0 +2\sum_{m=1}^Nc_m\cos(m\delta k z)\right)\le \frac{2}{\pi L}\,.
\end{equation}
If the inequality holds for $z=0$, it holds for all $z$. Then taking the inequality at $z=0$ the amplitude coefficients $c_m$ must satisfy
\begin{equation}
    \label{eq:restriction on c_n}
    c_0 +2\sum_{m=1}^N c_m \le \sqrt{\frac{2}{\pi}}\frac{1}{L\sigma_k}.
\end{equation}
If we ensure the prefactor coefficients satisfy the above inequality we guarantee that the target PMF amplitude can always be tracked by changing the domains of the crystal.  More sophisticated treatments can be made to determine the optimal choice of constants $c_m$ by considering different choices of $z$ but as a first considering we stop with the inequality in Eq. \eqref{eq:restriction on c_n}. 
\begin{figure}[h]
    \centering
    \includegraphics[width=0.45\columnwidth]{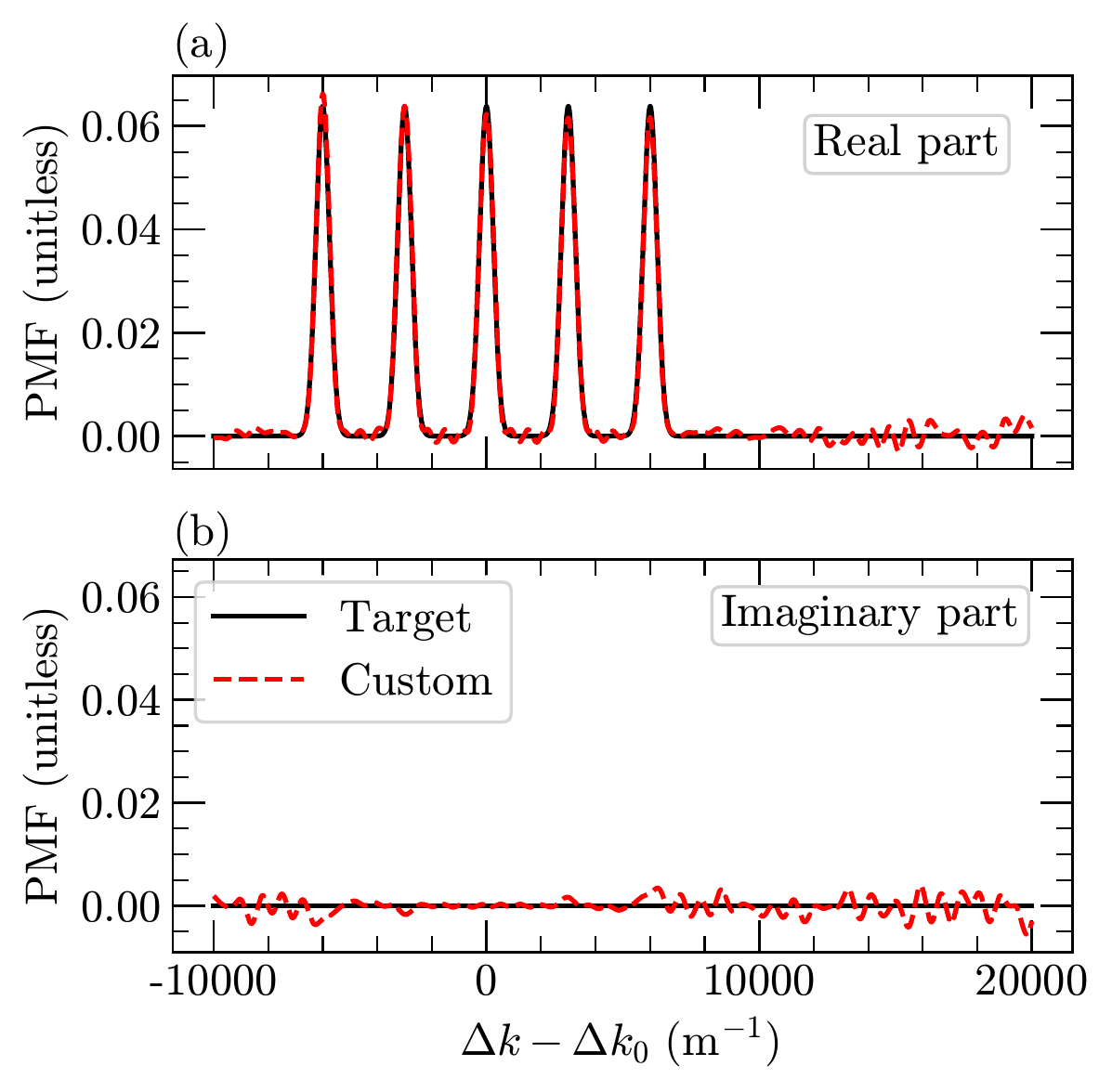}    \includegraphics[width=0.45\columnwidth]{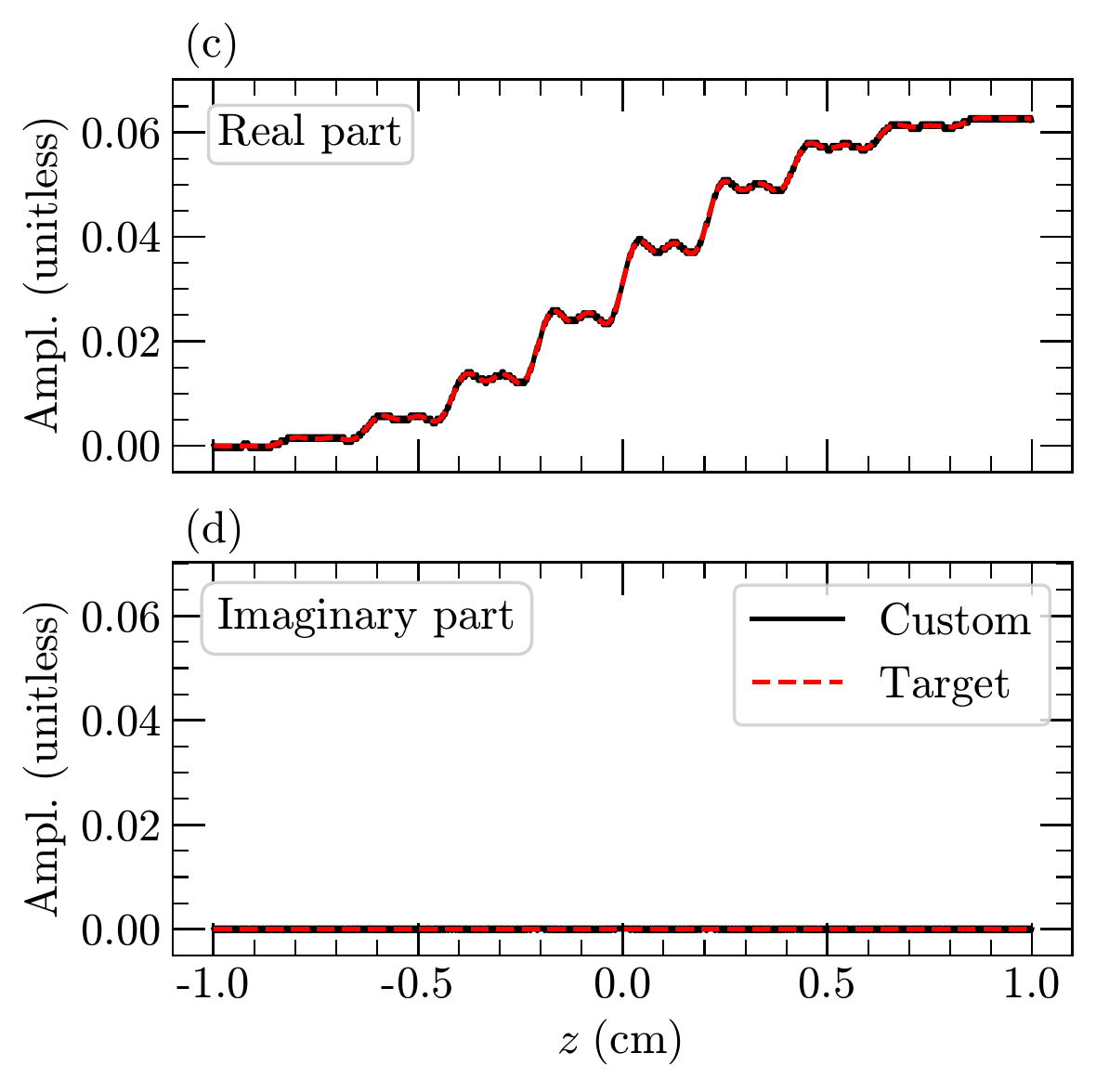}
    \caption{Real part (a) and imaginary part (b) of five-peak target phase-matching function as defined in Eq. \eqref{eq:target pmf} (with  $c_m = \sqrt{2}/(\sqrt{\pi}L\sigma_k 5)$, $\sigma_k=2.5/L$ and $\delta k=24\sigma_k$) and real part (c) and imaginary part (d) of corresponding amplitude function. We took $\sigma_k=2.5/L$ to ensure that the target nonlinearity profile fits within the length of the crystal. The custom-poled crystal has $N=1073$ domains of width $w=18.63\mu m$. The domain width was chosen to match the phase-matching conditions of Type-II KTP in the symmetric group-velocity-matched regime.}
    \label{fig:A1}
\end{figure}

\end{document}

%% file: Appendices.tex
\onecolumngrid

%% file: main.bbl
\begin{thebibliography}{10}

\bibitem{milburn1999quantum}
GJ~Milburn and Samuel~L Braunstein.
\newblock Quantum teleportation with squeezed vacuum states.
\newblock {\em Physical Review A}, 60(2):937, 1999.

\bibitem{hillery2000quantum}
Mark Hillery.
\newblock Quantum cryptography with squeezed states.
\newblock {\em Physical Review A}, 61(2):022309, 2000.

\bibitem{arrazola2021quantum}
JM~Arrazola, V~Bergholm, K~Br{\'a}dler, TR~Bromley, MJ~Collins, I~Dhand,
  A~Fumagalli, T~Gerrits, A~Goussev, LG~Helt, et~al.
\newblock Quantum circuits with many photons on a programmable nanophotonic
  chip.
\newblock {\em Nature}, 591(7848):54--60, 2021.

\bibitem{bennink2004quantum}
Ryan~S Bennink, Sean~J Bentley, Robert~W Boyd, and John~C Howell.
\newblock Quantum and classical coincidence imaging.
\newblock {\em Physical review letters}, 92(3):033601, 2004.

\bibitem{schnabel2010quantum}
Roman Schnabel, Nergis Mavalvala, David~E McClelland, and Ping~K Lam.
\newblock Quantum metrology for gravitational wave astronomy.
\newblock {\em Nature communications}, 1(1):1--10, 2010.

\bibitem{lvovsky2014squeezed}
A.~I. Lvovsky.
\newblock Squeezed light, 2014.

\bibitem{gessner2018sensitivity}
Manuel Gessner, Luca Pezz{\`e}, and Augusto Smerzi.
\newblock Sensitivity bounds for multiparameter quantum metrology.
\newblock {\em Physical review letters}, 121(13):130503, 2018.

\bibitem{sokolov2004squeezed}
Ivan~V Sokolov and Mikhail~I Kolobov.
\newblock Squeezed-light source for superresolving microscopy.
\newblock {\em Optics letters}, 29(7):703--705, 2004.

\bibitem{lian2007continuous}
Yimin Lian, Changde Xie, and Kunchi Peng.
\newblock Continuous variable multipartite entanglement and optical
  implementations of quantum communication networks.
\newblock {\em New Journal of Physics}, 9(9):314, 2007.

\bibitem{fabre2020generation}
Nicolas Fabre, G~Maltese, F~Appas, S~Felicetti, A~Ketterer, A~Keller,
  T~Coudreau, F~Baboux, MI~Amanti, S~Ducci, et~al.
\newblock Generation of a time-frequency grid state with integrated biphoton
  frequency combs.
\newblock {\em Physical Review A}, 102(1):012607, 2020.

\bibitem{Yeoman1993TwomodeSG}
Guy Yeoman and Stephen~Mark Barnett.
\newblock Two-mode squeezed gaussons.
\newblock 1993.

\bibitem{van2000multipartite}
Peter van Loock and Samuel~L Braunstein.
\newblock Multipartite entanglement for continuous variables: a quantum
  teleportation network.
\newblock {\em Physical Review Letters}, 84(15):3482, 2000.

\bibitem{elezov2018active}
MS~Elezov, ML~Scherbatenko, DV~Sych, and GN~Goltsman.
\newblock Active and passive phase stabilization for the all-fiber michelson
  interferometer.
\newblock In {\em Journal of Physics: Conference Series}, volume 1124, 2018.

\bibitem{pfister2004multipartite}
Olivier Pfister, Sheng Feng, Gregory Jennings, Raphael Pooser, and Daruo Xie.
\newblock Multipartite continuous-variable entanglement from concurrent
  nonlinearities.
\newblock {\em Physical Review A}, 70(2):020302, 2004.

\bibitem{menicucci2007ultracompact}
Nicolas~C Menicucci, Steven~T Flammia, Hussain Zaidi, and Olivier Pfister.
\newblock Ultracompact generation of continuous-variable cluster states.
\newblock {\em Physical Review A}, 76(1):010302, 2007.

\bibitem{alexander2016one}
Rafael~N Alexander, Pei Wang, Niranjan Sridhar, Moran Chen, Olivier Pfister,
  and Nicolas~C Menicucci.
\newblock One-way quantum computing with arbitrarily large time-frequency
  continuous-variable cluster states from a single optical parametric
  oscillator.
\newblock {\em Physical Review A}, 94(3):032327, 2016.

\bibitem{zhu2021hypercubic}
Xuan Zhu, Chun-Hung Chang, Carlos Gonz{\'a}lez-Arciniegas, Avi Pe’er, Jacob
  Higgins, and Olivier Pfister.
\newblock Hypercubic cluster states in the phase-modulated quantum optical
  frequency comb.
\newblock {\em Optica}, 8(3):281--290, 2021.

\bibitem{ansari2018tailoring}
Vahid Ansari, John~M Donohue, Benjamin Brecht, and Christine Silberhorn.
\newblock Tailoring nonlinear processes for quantum optics with pulsed
  temporal-mode encodings.
\newblock {\em Optica}, 5(5):534--550, 2018.

\bibitem{morrison2022frequency}
Christopher~L Morrison, Francesco Graffitti, Peter Barrow, Alexander Pickston,
  Joseph Ho, and Alessandro Fedrizzi.
\newblock Frequency-bin entanglement from domain-engineered down-conversion.
\newblock {\em arXiv preprint arXiv:2201.07259}, 2022.

\bibitem{lloyd1993programming}
Seth Lloyd.
\newblock Programming pulse driven quantum computers.
\newblock {\em Science}, 261(quant-ph/9912086):1569, 1993.

\bibitem{couteau2018spontaneous}
Christophe Couteau.
\newblock Spontaneous parametric down-conversion.
\newblock {\em Contemporary Physics}, 59(3):291--304, 2018.

\bibitem{weiner2011ultrafast}
Andrew~M Weiner.
\newblock Ultrafast optical pulse shaping: A tutorial review.
\newblock {\em Optics Communications}, 284(15):3669--3692, 2011.

\bibitem{froehly1983ii}
Cl~Froehly, B~Colombeau, and M~Vampouille.
\newblock Ii shaping and analysis of picosecond light pulses.
\newblock In {\em Progress in optics}, volume~20, pages 63--153. Elsevier,
  1983.

\bibitem{monmayrant2010newcomer}
Antoine Monmayrant, S{\'e}bastien Weber, and B{\'e}atrice Chatel.
\newblock A newcomer's guide to ultrashort pulse shaping and characterization.
\newblock {\em Journal of Physics B: Atomic, Molecular and Optical Physics},
  43(10):103001, 2010.

\bibitem{quesada2018gaussian}
Nicol{\'a}s Quesada and Agata~M Bra{\'n}czyk.
\newblock Gaussian functions are optimal for waveguided
  nonlinear-quantum-optical processes.
\newblock {\em Physical Review A}, 98(4):043813, 2018.

\bibitem{boyd2019nonlinear}
Robert~W Boyd.
\newblock {\em Nonlinear optics}.
\newblock Academic press, 2019.

\bibitem{zukauskas2011fabrication}
Andrius Zukauskas, Gustav Str{\"o}mqvist, Valdas Pasiskevicius, Fredrik
  Laurell, Michael Fokine, and Carlota Canalias.
\newblock Fabrication of submicrometer quasi-phase-matched devices in ktp and
  rktp.
\newblock {\em Optical Materials Express}, 1(7):1319--1325, 2011.

\bibitem{fejer1992quasi}
Martin~M Fejer, GA~Magel, Dieter~H Jundt, and Robert~L Byer.
\newblock Quasi-phase-matched second harmonic generation: tuning and
  tolerances.
\newblock {\em IEEE Journal of Quantum Electronics}, 28(11):2631--2654, 1992.

\bibitem{imeshev2000ultrashort}
G~Imeshev, MA~Arbore, MM~Fejer, A~Galvanauskas, M~Fermann, and D~Harter.
\newblock Ultrashort-pulse second-harmonic generation with longitudinally
  nonuniform quasi-phase-matching gratings: pulse compression and shaping.
\newblock {\em JOSA B}, 17(2):304--318, 2000.

\bibitem{arbore1997engineerable}
MA~Arbore, A~Galvanauskas, D~Harter, MH~Chou, and MM~Fejer.
\newblock Engineerable compression of ultrashort pulses by use of
  second-harmonic generation in chirped-period-poled lithium niobate.
\newblock {\em Optics Letters}, 22(17):1341--1343, 1997.

\bibitem{Dixon2013}
P.~Ben Dixon, Jeffrey~H. Shapiro, and Franco N.~C. Wong.
\newblock Spectral engineering by gaussian phase-matching for quantum
  photonics.
\newblock {\em Opt. Express}, 21(5):5879--5890, Mar 2013.

\bibitem{chen2017efficient}
Changchen Chen, Cao Bo, Murphy~Yuezhen Niu, Feihu Xu, Zheshen Zhang, Jeffrey~H
  Shapiro, and Franco~NC Wong.
\newblock Efficient generation and characterization of spectrally factorable
  biphotons.
\newblock {\em Optics express}, 25(7):7300--7312, 2017.

\bibitem{chen2019indistinguishable}
Changchen Chen, Jane~E Heyes, Kyung-Han Hong, Murphy~Yuezhen Niu, Adriana~E
  Lita, Thomas Gerrits, Sae~Woo Nam, Jeffrey~H Shapiro, and Franco~NC Wong.
\newblock Indistinguishable single-mode photons from spectrally engineered
  biphotons.
\newblock {\em Optics express}, 27(8):11626--11634, 2019.

\bibitem{PhysRevApplied.12.034059}
Chaohan Cui, Reeshad Arian, Saikat Guha, N.~Peyghambarian, Quntao Zhuang, and
  Zheshen Zhang.
\newblock Wave-function engineering for spectrally uncorrelated biphotons in
  the telecommunication band based on a machine-learning framework.
\newblock {\em Phys. Rev. Applied}, 12:034059, Sep 2019.

\bibitem{branczyk2011engineered}
Agata~M Bra{\'n}czyk, Alessandro Fedrizzi, Thomas~M Stace, Tim~C Ralph, and
  Andrew~G White.
\newblock Engineered optical nonlinearity for quantum light sources.
\newblock {\em Optics express}, 19(1):55--65, 2011.

\bibitem{tambasco2016domain}
JL~Tambasco, A~Boes, LG~Helt, MJ~Steel, and A~Mitchell.
\newblock Domain engineering algorithm for practical and effective photon
  sources.
\newblock {\em Optics express}, 24(17):19616--19626, 2016.

\bibitem{dosseva2016shaping}
Annamaria Dosseva, {\L}ukasz Cincio, and Agata~M Bra{\'n}czyk.
\newblock Shaping the joint spectrum of down-converted photons through
  optimized custom poling.
\newblock {\em Physical Review A}, 93(1):013801, 2016.

\bibitem{graffitti2018independent}
Francesco Graffitti, Peter Barrow, Massimiliano Proietti, Dmytro Kundys, and
  Alessandro Fedrizzi.
\newblock Independent high-purity photons created in domain-engineered
  crystals.
\newblock {\em Optica}, 5(5):514--517, 2018.

\bibitem{graffitti2017pure}
Francesco Graffitti, Dmytro Kundys, Derryck~T Reid, Agata~M Bra{\'n}czyk, and
  Alessandro Fedrizzi.
\newblock Pure down-conversion photons through sub-coherence-length domain
  engineering.
\newblock {\em Quantum Science and Technology}, 2(3):035001, 2017.

\bibitem{pickston2021optimised}
Alexander Pickston, Francesco Graffitti, Peter Barrow, Christopher Morrison,
  Joseph Ho, Agata~M Bra{\'n}czyk, and Alessandro Fedrizzi.
\newblock Optimised domain-engineered crystals for pure telecom photon sources.
\newblock {\em arXiv preprint arXiv:2101.08280}, 2021.

\bibitem{graffitti2018design}
Francesco Graffitti, J{\'e}r{\'e}my Kelly-Massicotte, Alessandro Fedrizzi, and
  Agata~M Bra{\'n}czyk.
\newblock Design considerations for high-purity heralded single-photon sources.
\newblock {\em Physical Review A}, 98(5):053811, 2018.

\bibitem{yamada1993first}
M~Yamada, N~Nada, M~Saitoh, and K~Watanabe.
\newblock First-order quasi-phase matched linbo3 waveguide periodically poled
  by applying an external field for efficient blue second-harmonic generation.
\newblock {\em Applied Physics Letters}, 62(5):435--436, 1993.

\bibitem{quesada2017you}
Nicol{\'a}s Quesada and JE~Sipe.
\newblock Why you should not use the electric field to quantize in nonlinear
  optics.
\newblock {\em Optics letters}, 42(17):3443--3446, 2017.

\bibitem{helt2017parasitic}
LG~Helt, Agata~M Bra{\'n}czyk, Marco Liscidini, and MJ~Steel.
\newblock Parasitic photon-pair suppression via photonic stop-band engineering.
\newblock {\em Physical Review Letters}, 118(7):073603, 2017.

\bibitem{scully_zubairy_1997}
Marlan~O. Scully and M.~Suhail Zubairy.
\newblock {\em Quantum Optics}.
\newblock Cambridge University Press, 1997.

\bibitem{lawrie2019quantum}
Benjamin~J Lawrie, Paul~D Lett, Alberto~M Marino, and Raphael~C Pooser.
\newblock Quantum sensing with squeezed light.
\newblock {\em ACS Photonics}, 6(6):1307--1318, 2019.

\bibitem{aasi2013enhanced}
Junaid Aasi, J~Abadie, BP~Abbott, Richard Abbott, TD~Abbott, MR~Abernathy, Carl
  Adams, Thomas Adams, Paolo Addesso, RX~Adhikari, et~al.
\newblock Enhanced sensitivity of the ligo gravitational wave detector by using
  squeezed states of light.
\newblock {\em Nature Photonics}, 7(8):613--619, 2013.

\bibitem{wang2007quantum}
Xiang-Bin Wang, Tohya Hiroshima, Akihisa Tomita, and Masahito Hayashi.
\newblock Quantum information with gaussian states.
\newblock {\em Physics reports}, 448(1-4):1--111, 2007.

\bibitem{armstrong2012programmable}
Seiji Armstrong, Jean-Fran{\c{c}}ois Morizur, Jiri Janousek, Boris Hage,
  Nicolas Treps, Ping~Koy Lam, and Hans-A Bachor.
\newblock Programmable multimode quantum networks.
\newblock {\em Nature communications}, 3(1):1--8, 2012.

\bibitem{epping2017multi}
Michael Epping, Hermann Kampermann, Dagmar Bru{\ss}, et~al.
\newblock Multi-partite entanglement can speed up quantum key distribution in
  networks.
\newblock {\em New Journal of Physics}, 19(9):093012, 2017.

\bibitem{yang2008spontaneous}
Zhenshan Yang, Marco Liscidini, and JE~Sipe.
\newblock Spontaneous parametric down-conversion in waveguides: A backward
  heisenberg picture approach.
\newblock {\em Physical Review A}, 77(3):033808, 2008.

\bibitem{shankar2012principles}
Ramamurti Shankar.
\newblock {\em Principles of quantum mechanics}.
\newblock Springer Science \& Business Media, 2012.

\end{thebibliography}
